\documentclass[12pt]{iopart}

\usepackage{hyperref,iopams,amsfonts}

\usepackage{graphicx}

\newcommand{\DOIURL}[1]{http://dx.doi.org/#1}
\newcommand{\PhysRL}[2]%
    {\href{\DOIURL{10.1103/PhysRevLett.#1.#2}}%
    {Phys. Rev. Lett. \textbf{#1}, #2}}
\newcommand{\PRA}[2]%
    {\href{\DOIURL{10.1103/PhysRevA.#1.#2}}%
    {Phys. Rev. A \textbf{#1}, #2}}
\newcommand{\RevMP}[2]%
    {\href{\DOIURL{10.1103/RevModPhys.#1.#2}}%
    {Rev. Mod. Phys. \textbf{#1}, #2}}
\newcommand{\QPH}[1]%
    {\href{http://arxiv.org/abs/quant-ph/#1}%
    {\texttt{quant-ph/#1}}}

\newcommand{\ie}{\emph{i.e. }}
\newcommand{\Id}{\mathbb I}
\newcommand\odelta{\delta\!}

\begin{document}

\title{Feasibility of a quantum memory for continuous variables
based on trapped ions}

\author{Thomas Coudreau$^1$, Fr\'{e}d\'{e}ric Grosshans$^2$, Samuel Guibal$^1$,
  Luca Guidoni$^1$}  
\ead{\href{mailto:coudreau@spectro.jussieu.fr}{coudreau@spectro.jussieu.fr}}

\address{$^1$Laboratoire Mat\'{e}riaux et Ph\'{e}nom\`{e}nes Quantiques, CNRS
UMR 7162, Case 7021, Universit\'{e} Denis Diderot, 2 Place Jussieu,
75251 Paris cedex 05, France}

\address{$^2$Laboratoire de Photonique Quantique et Moléculaire, CNRS
UMR 8537, Ecole Normale Supérieure de Cachan, 61 avenue du Président Wilson,
94235 Cachan, France}

\date{\today}

\begin{abstract}
We propose to use a large cloud of cold trapped ions as a
medium for quantum optics and quantum information experiments.
Contrary to most recent realizations of qubit manipulation based on
a small number of trapped and cooled ions, we study the case of
traps containing a macroscopic number of ions. We consider in
particular the implementation of a quantum memory for quantum
information stored in continuous variables and study the impact of
the relevant physical parameters on the expected performances of the
system.\end{abstract}


\vspace{2pc}

\maketitle

\section{Introduction}

As quantum information becomes closer to applications, the need
for quantum memories becomes stronger
\cite{divincenzo,dlcz,reviewlukin}. Quantum memories are systems
which store in so-called "static qubits" the  quantum information
usually carried by "flying qubits", namely photons. A single ion
or a pair of ions forms a prototype of such a quantum memory
\cite{memory_wineland,memory_monroe,memory_wineland2}. It has also
been shown that an ensemble of atoms can successfully store the
quantum information carried by a single photon
\cite{dlcz,kuzmich,kimble}. However, quantum information
processing is not limited to discrete variables: continuous
variable systems also form an attractive tool for such a purpose
\cite{CV,brareview,cvqipbook}. In this context, the use of an
ensemble of atoms as a quantum memory has been proposed
\cite{propmemory,pinard} and experimentally demonstrated
\cite{polzik}. In this realization, both the storage medium and
the light beam are described by continuous variables.

The goal of this paper is to study the feasibility of a quantum
memory relying on a large cloud of cold trapped ions. As we will
show, such a system has several advantages with respect to the
neutral atom clouds. Among those, one can mention the strong
confinement of the ions that opens the way to an efficient
interaction with light while keeping the sample immobile in the interaction
region for a virtually infinite time.
The paper is organized as follows: we will start
by describing the basic requisites for a quantum memory and by
briefly recalling the principles which govern a light-matter
interface suitable for such a purpose. Then, we will propose an
experimental implementation based on a cloud of cold trapped ions.
We will finally present the expected properties with respect to
decoherence and memory lifetime of such a system.

\section{Continuous Variable and Quantum Memory Principles}

In the case considered here of a quantum memory for continuous
variables, quantum information is carried by the quadrature
components of a pulse of light such as in
\cite{polzik,pinard,pinard_simple}. We will quickly recall the
continuous variable formalism in subsection~\ref{secCV}, and then
introduce the criteria that define the boundaries between the
classical and quantum regimes for a quantum memory in
subsection~\ref{secQmemCriter}.

\subsection{Continuous Variables}
\label{secCV}
\subsubsection{Continuous Observables}
The observables studied in Continuous Variable quantum systems
\cite{AnnPhys,Symplectic,EisertWolf} (the quadratures) are
formally analogue to the position and momentum of a particle and
hence usually noted $\hat Q$ and $\hat P$. In the system described
here, the ``quadratures'' of the ion cloud will be the projection
of the global spin of the cloud along some direction, while the
corresponding variables of the light beam will be two Stokes
operators describing its polarization \cite{duan00,polzikEPR}.
When suitably normalized, they obey the canonical commutation
relation\cite{AnnPhys,Symplectic}
\begin{equation}
    \label{commut}
    [\hat Q, \hat P]=2i.
\end{equation}

The above commutation relation implies the Heisenberg relation
\begin{equation}
\odelta Q\,\odelta P \ge 1,
\end{equation}
where $\odelta Q$ (resp. $\odelta P$) is the standard deviation
of $Q$ (resp. $P$).
The fact that the quadratures can take any real value also comes from this
commutation relation.
%

\subsubsection{Gaussian States and Channels}

When the characteristic functions and quasiprobability distributions
of the system are Gaussian, it is said to be in a Gaussian state. If
Gaussian states are the simplest to study theoretically, they are
also fortunately the most common in experiments. We will therefore
restrict ourselves to Gaussian states in this article. Since the
probability distributions associated with the quadratures of a
Gaussian state are Gaussian, such a state is fully characterized by
the average value and the covariance matrix of the continuous
observables.

A Gaussian operation ---\ie\ an operation which preserves the
Gaussian character of the state--- can then be described by a
linear action on the quadratures, possibly including the coupling
with an external mode. For example, the action of a lossy channel
with absorption $\varepsilon$ on a mode $(\hat Q_{\mathrm{in}},\hat
P_{\mathrm{in}})$ is given by
\begin{eqnarray}
    \hat Q_{\mathrm{out}}&=\sqrt{1-\varepsilon}\hat Q_{\mathrm{in}}
                       + \sqrt{\varepsilon}\hat Q_{\mathrm{vac}}\\
 \hat P_{\mathrm{out}}&=\sqrt{1-\varepsilon}\hat P_{\mathrm{in}}
                       + \sqrt{\varepsilon}\hat P_{\mathrm{vac}},
\label{eqAbschannel}\end{eqnarray}
where $(\hat Q_{\mathrm{vac}},\hat P_{\mathrm{vac}})$ is a mode in a
vacuum state \ie $\hat Q_{\mathrm{vac}}$ and $\hat P_{\mathrm{vac}}$ are
uncorrelated to the input mode, and have zero average value and unit
variance.

\subsection{Criteria for quantum memory}
\label{secQmemCriter}

\subsubsection{Generic criteria}
A quantum memory, by definition, stores informations about a quantum
state for a given time interval, and it should do it better than any
classical memory (\ie classical-states based memory). Since an
\emph{a priori} known quantum state has a complete classical
description (its density matrix), it can be reconstructed with an
arbitrarily high fidelity by a setup only storing this description
in a classical memory. Therefore, similarly to quantum teleportation
criteria \cite{criteres_teleport}, a quantum memory criterion can
only be defined when the input is unknown, \ie when the input state
is drawn at random from a given set $\mathcal S$ of possible states.

The output of the quantum memory can either be a quantum state or
classical information. 
In the following, we will consider these two situations that we
name the \emph{Identity Quantum Memory} (IdQM) and the
\emph{Delayed Measurement Quantum Memory} (DMQM). The output of an
\emph{Identity Quantum Memory} should be a quantum state as close
as possible to the input state. The input-output mapping of a
perfect IdQM is then the identity operator, hence its name.
Applications of IdQMs include storing intermediate states in
quantum computation and the realization of collective attacks
against quantum cryptography \cite{DevetakWinter}.
The best classical setup 
devoted to the task of an IdQM, first measures the input state, then
stores the obtained information in a classical memory and finally
tries to reconstruct the input state from this information. This
reconstruction problem has been well studied and the best classical
reconstruction fidelity is well known in a lot of situations
\cite{criteres_teleport,NGClone}.
The usual figure of merit for such a memory is the (average or
minimal) fidelity, but any distortion measure can be used. Among
them, entropic quantities suit to cryptographic applications
\cite{DevetakWinter}.

While a perfect IdQM allows to construct any conceivable quantum
memory, the corresponding criterion, defining the \emph{worst}
IdQM, is not always relevant, as shown below.

Another quantum memory application commonly discussed in the
literature is related to the delayed measurement problem
\cite{DelayedChoice}.
A delayed measurement quantum memory (DMQM) receives a quantum state
to be stored, and only later learns the measurement to be performed
on that state. The output of the DMQM is the classical result of
this target-measurement. A commonly discussed application of a DMQM
is the realization of individual attacks against quantum
cryptographic schemes. As above, the suitable figure of merit
changes with the memory application, but, in any case, the fidelity
is not the relevant figure of merit, since the output of this
quantum memory is classical. The optimal classical strategy to
perform a classical measurement is to immediately measure the
quantum state and store the result in a classical memory. When the
target-measurement is drawn from a set of incompatible measurements,
the best classical measurement will be approximative, since it has
to guess the result of all incompatible target-measurement. However,
this optimal classical strategy doesn't need to reconstruct the
input state, which usually renders the criterion for DMQM stricter
than the IdQM one. In other words, if an IdQM followed by the
relevant measurement is a possible way to build a DMQM, the
resulting DMQM is not guaranteed to be better than its optimal
classical counterpart, even if the IdQM is above the quantum
threshold.

The two definition above can be formalized in a more general
framework, described below. A quantum memory is a channel taking
two inputs, one of which is a quantum state drawn from a known set
$\mathcal S=\{\rho_i\}_i$  with probability $p_i$, and the other
is an (optional) delayed information $\tau$. The action of the
memory on $\rho$ can be described by the completely positive map
$\phi_\tau$, which should be as close as possible to a target map
$\phi_\tau^{\mathrm{target}}$. The target map corresponding to the
IdQM described above is the identity $\phi_\tau^{\mathrm{target}} =
\Id$ and doesn't depend on $\tau$. For the DMQM, the
$\phi_\tau^{\mathrm{target}}$ correspond to the various
measurements, each $\tau$ identifying a particular measurement.

The optimal classical strategy trying to simulate
$\phi_\tau^{\mathrm{target}}$ defines the criterion for a quantum
memory. This strategy should be based on measurements, classical
memory and state reconstruction. Furthermore, the measurements
must not depend on $\tau$.

One can see from the above examples and definitions that there are
as many criteria for quantum memories as there are applications,
the criteria depending on the input state set $\mathcal S$, the
target map $\phi_\tau$,  and the chosen figure of merit.

\subsubsection{Continuous Variable quantum memories}
\label{sec:cvqm}

As mentioned before, we will restrict ourselves to continuous
variables-quantum memories (IdQM and DMQM). Furthermore, we will
restrict ourself to gaussian quantum memories\footnote{even if
they are not always optimal for commonly used figures of merit
\cite{NGClone}}. Therefore $\mathcal S$ is a covariant set of
gaussian states with uniform distribution in phase-space.

Since the quantum memory is gaussian, it can be described like a
Gaussian Channel, with the following input-output relation :
\begin{eqnarray}
    \hat Q_{\mathrm{out}}&=\sqrt{G_Q}\hat Q_{\mathrm{in}}
                       + \sqrt{N_Q}\hat Q_{\mathrm{vac}}\\
 \hat P_{\mathrm{out}}&=\sqrt{G_P}\hat P_{\mathrm{in}}
                       \pm \sqrt{N_P}\hat P_{\mathrm{vac}},
\end{eqnarray}
with $N\ge  |G-1|$, where the noise $N=\sqrt{N_QN_P}$
and the gain $G=\sqrt{G_QG_P}$.

In the following, we will describe the criteria for IdQM and DMQM,
illustrating them \emph{via} the example of an optical fiber loop.

The best classical IdQM approximation is well studied since the
early days of continuous variable teleportation
\cite{criteres_teleport}, and has a unity gain and a noise
equivalent to twice the shot noise on both quadratures. An IdQM
should therefore verify
\begin{equation}
   G_Q=G_P=1  \qquad N<2. \label{eq:crit_idqm}
\end{equation}
Note that if $G_Q\neq G_P$ but $G=1$, the memory squeezes the
input. Since the squeezing is a reversible operation, such a
memory can be considered as a generalized IdQM.

As easily seen from eq. \ref{eqAbschannel},the noise added by an
optical fiber of transmission $T=1-\varepsilon$ has a variance
equal to $\varepsilon=1-T$. If one wants to use this fiber as an
IdQM, one needs to correct for the transmission gain $T$ by
preamplifying the input with a gain $1/T$, adding a noise of at
least $1/T -1$. The resulting channel has then the wanted unit
gain, and a noise $N=2(1-T)$, always smaller than the classical
limit of $2$. However, $T$ decreases exponentially, at a rate of
0.04 $\mathrm{dB}/\mu\mathrm{s}$ in a fiber of refracting index
$n=1.5$ and attenuation 0.2 $\mathrm{dB}/\mathrm{km}$. The
transmission $T$ being divided by 100 every 500 $\mu\mathrm s$,
this memory quickly becomes indistinguishable from a classical
memory, even in the absence of technical noise.

The canonical continuous variable DMQM allows to measure either
$Q$ or $P$. The corresponding optimal classical measurement has
been studied since Heisenberg, and the noise added by this
measurement obeys the uncertainty relationship $\odelta Q \cdot
\odelta P \ge 1$. The violation of this relationship, written
\begin{equation}
  \frac NG < 1 \label{eq:crit_dmqm}
\end{equation}
is therefore a
criterion for continuous variable DMQM.

If one implements a DMQM
with a fiber loop followed by a (perfect) homodyne measurement,
the relevant figure of merit is the equivalent input noise $N/G$
\cite{AnnPhys}
of the fiber, and the criterion can be written
\begin{equation*}
  \frac1T - 1 < 1
   \Leftrightarrow
  T>\frac12,
\end{equation*}
which corresponds to a maximal storage time of 75~$\mu$s in an
optical fiber loop.

\subsection{Quantum memory with an X-type level scheme}
\label{sec:principes}

\subsubsection{Light-matter interaction}

\begin{figure}[h]
\centerline{\includegraphics[width=.75\columnwidth]{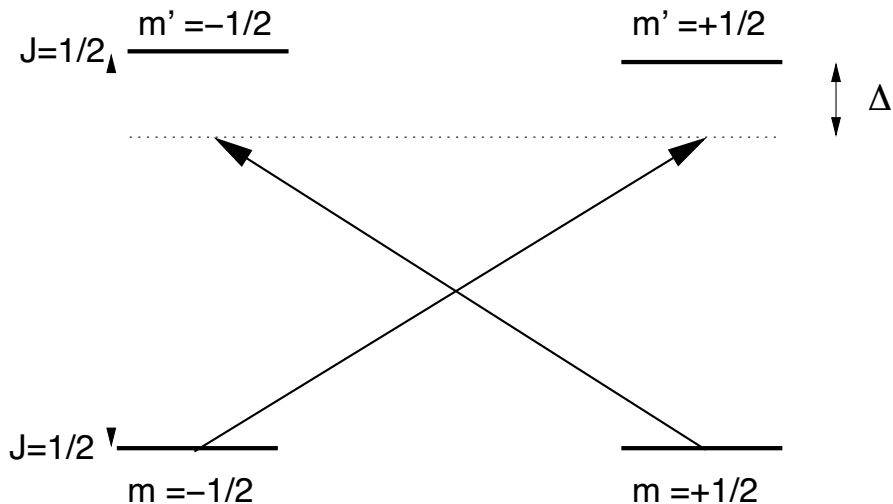}}
\caption{\label{fig:niveaux_Sr_info} X-type level scheme of the
atomic medium used in the quantum memory. The laser is detuned by
$\Delta$ with respect to a 1/2 \ensuremath{\to} 1/2 transition
characterized by a pulsation $\omega$ and excited-state lifetime
$1/\gamma$.}
\end{figure}

The unique experimental realization of a continuous variable quantum
memory to date is based on a three step process \cite{polzik}:
first, the light pulse carrying the quantum information is sent
through the sample; the output beam is then measured; and finally a
feed-back is performed on the sample conditioned on this
measurement. In this realization, the levels involved are considered
as an X-type system, with two levels in the ground state and two in
the excited state coupled by two lasers
(Fig.~\ref{fig:niveaux_Sr_info}). Other level schemes have also been
proposed. For instance, when a $\mathrm{\Lambda}$-type level scheme
is used (with only one excited state and two lasers), a single
interaction is sufficient \cite{pinard,pinard_simple}. Other
proposed schemes, involve multiple passages through the atomic
medium \cite{multipass} and no back-action \emph{via} a magnetic
field pulse. In this paper, we will focus on the first scheme.

We will follow here the formalism introduced in \cite{duan00}. Let
us begin by recalling its main properties. The medium consists in
$N_a$ atoms with a 1/2 \ensuremath{\to} 1/2 transition coupled to
a laser beam close to resonance (figure
\ref{fig:niveaux_Sr_info}). We consider that the information is
carried in a pulse with duration $T$, much larger than the atomic
response time that propagates along the $z$ direction. We denote
$N_p$ the total number of photons contained in the pulse.

The basic idea is that the quantum information stored in the
polarization state described by quantum Stokes operators is
coupled \emph{via} the Faraday effect to the spin operators that
characterize the collective spin of the atomic medium.

The collective spin operators of the atomic medium read
\begin{equation}
\hat J_{x,y,z} = \sum_{i=1}^{N_a} \hat \sigma_{x,y,z}^{(i)}
\end{equation}
where $\hat \sigma_{x,y,z}^{(i)}$ denotes the individual spin of
atom $i$. The light field is described by its quantum Stokes
operators \cite{polar}:
\begin{eqnarray}
\hat S_0 &=& \hat a^\dagger_x \hat a_x + \hat a^\dagger_y \hat
a_y\nonumber\\
\hat S_1 &=& \hat a^\dagger_x \hat a_x - \hat a^\dagger_y \hat
a_y\nonumber\\
\hat S_2 &=& \hat a^\dagger_x \hat a_y + \hat a^\dagger_y \hat
a_x= \hat a^\dagger_{+45^\circ} \hat a_{+45^\circ} + \hat
a^\dagger_{- 45^\circ} \hat a_{- 45^\circ}\nonumber\\
\hat S_3 &=& i \left(\hat a^\dagger_y \hat a_x - \hat a^\dagger_x
\hat a_y \right) = \hat a^\dagger_{\sigma^+} \hat a_{\sigma^+} +
\hat a^\dagger_{\sigma^-} \hat a_{\sigma^-}
\end{eqnarray}
where $\hat a_{x,y,\pm 45^\circ,\sigma^\pm}$ and $\hat a_{x,y,\pm
45^\circ,\sigma^\pm}^\dagger$ denote respectively the annihilation
and creation operators in the horizontal, vertical, $\pm 45^\circ$,
left- and right-circular polarizations. The $\hat S_i$ and $\hat
J_k$ operators verify the usual commutation relations:
\begin{eqnarray}
\left[\hat S_1 ,\hat S_2\right] = i \hat S_3 \quad && \quad
\left[\hat J_x ,\hat J_y\right] = i \hat J_z,\\
\left[ \hat S_2 ,\hat S_3\right]= i \hat S_1 \quad && \quad
\left[\hat J_y ,\hat J_z\right] = i \hat J_x ,\\
\left[\hat S_3,\hat S_1\right] = i \hat S_2 \quad && \quad
\left[\hat J_z ,\hat J_x\right] = i \hat J_y.
\end{eqnarray}

We consider an input beam linearly polarized along the horizontal
direction, so that \[ \langle \hat S_1 \rangle = N_p, \langle \hat
S_2 \rangle = 0 = \langle \hat S_3 \rangle.
\]
We likewise consider an atomic cloud initially polarized in the
$x$ direction (for instance by optical pumping), so that
\[
\langle \hat J_x \rangle =N_a, \langle \hat J_y \rangle = 0 =
\langle \hat J_z \rangle.
\]
One can then define two pairs of operators associated with the
light beam and with the atomic sample,
\begin{eqnarray*}
\hat Q_p = \frac{\hat S_2}{\sqrt{\langle \hat S_1\rangle/2}}
&\mbox{\quad and \quad}&  \hat P_p = \frac{\hat S_3}{\sqrt{\langle
\hat S_1\rangle/2}}, \\
\hat Q_a = \frac{\hat J_y}{\sqrt{\langle \hat J_x\rangle/2}}
&\mbox{\quad and \quad}&  \hat P_a = \frac{\hat J_z}{\sqrt{\langle
\hat J_x\rangle/2}}.
\end{eqnarray*}
If the interaction is weak, one can treat $\hat
S_1$ and $\hat J_x$ classically \cite{duan00} so that the above operators
 verify
the canonical commutation relations
\[[\hat Q_p, \hat
P_p] = 2i \mbox{\quad and \quad} [\hat Q_a, \hat P_a] = 2i.\]

If we consider a non resonant interaction ($\Delta \gg \gamma$),
an input-output relation through the medium can be written as
\cite{duan00}
\begin{eqnarray}
\hat Q_p^{(out)} &=& \sqrt{1 - \varepsilon_p} \left(\hat
Q_p^{(in)} + \kappa \hat P_a^{(in)} \right) + \sqrt{\varepsilon_p}
\hat Q_p^{(vac)} \label{eq:xpout}\\
\hat  Q_a^{(out)} &=& \sqrt{1 - \varepsilon_a} \left(\hat Q_a^{(in)}
+ \kappa \hat P_p^{(in)} \right) + \sqrt{\varepsilon_a}
\hat Q_a^{(vac)}\\
\hat P_p^{(out)} &=& \sqrt{1 - \varepsilon_p} \hat P_p^{(in)} +
\sqrt{\varepsilon_p} \hat P_p^{(vac)}\\
\hat P_a^{(out)} &=& \sqrt{1 - \varepsilon_a} \hat P_a^{(in)} +
\sqrt{\varepsilon_a} \hat P_a^{(vac)}\label{eq:pdout}
\end{eqnarray}
where
\[
\varepsilon_a = N_p |g|^2 \frac{\gamma}{\Delta^2} \mbox{~and~}
\varepsilon_p = N_a |g|^2 \frac{\gamma}{\Delta^2},
\]
can be interpreted as the atomic and photonic losses,
\[
\kappa = 2 \sqrt{N_p N_a} \frac{|g|^2}{\Delta},
\]
is the beam-sample coupling constant, $Q_{a,p}^{(vac)}$ and
$P_{a,p}^{(vac)}$ are entering noise with a shot-noise limited
variance; the atom-photon coupling constant is given by
\[
|g|^2 = \frac{3 c \lambda^2}{16 \pi^2 A} \gamma \] where $A$  is the
beam cross section and $\lambda$ the transition wavelength.

After such an interaction, the $\hat P_p$ quadrature of light has
been mapped onto the $\hat Q_a$ quadrature of the cloud (see
Eq.~\ref{eq:xpout}). A second operation is now necessary to map the
other quadrature of light ($Q_p$) onto the other quadrature of the
collective spin ($P_a$). As stated previously, this is done using a
measurement of the output light beam and a feed back onto the
memory.

\subsubsection{Measurement and feed-back}

Using standard quantum optics techniques, one can easily measure the
$\hat Q_p$ quadrature of the light beam emerging from the cloud.
This information can then be mapped onto the atomic operator $\hat
P_a$ using for example a radio-frequency magnetic field pulse
\cite{polzik}. This can be done using a gain $\sqrt \mathcal{G}$
which can be fixed arbitrarily. One has then
\begin{eqnarray*}
\hat P_a^{mem} &=& \hat P_a^{out} + \sqrt \mathcal{G} \hat Q_p^{out} \nonumber\\
&=& \sqrt{1 - \varepsilon_a} (1 + \sqrt \mathcal{G} \kappa)
\hat P_a^{(in)} + \sqrt \mathcal{G} \sqrt{1-\varepsilon_p} \hat Q_p^{(in)} +
 \nonumber\\
&& \qquad \left( \sqrt{\varepsilon_a} \hat P_a^{(vac)} + \sqrt
\mathcal{G} \sqrt{\varepsilon_p} \hat Q_p^{(vac)}\right)
\end{eqnarray*}

In order to cancel the atomic input noise, $\hat P_a^{(in)}$, one
can fix $\sqrt \mathcal{G} \kappa = -1$ which yields
\begin{equation}
\hat P_a^{mem} = - \frac{\sqrt{1-\varepsilon_p}}{\kappa} \hat
Q_p^{(in)} + \hat P_a^{vac \prime} \label{eq:pamem}
\end{equation}
with
\[
\hat P_a^{vac \prime} =\sqrt{\varepsilon_a} \hat P_a^{(vac)} -
\frac{ \sqrt{\varepsilon_p}}{\kappa} \hat Q_p^{(vac)}.
\]

Through these successive operations, the atomic medium quadratures
$(\hat Q_a, \hat P_a)$ contain a term proportional to the input
light quadratures $(\hat Q_l, \hat P_l)$. We have thus realized a
quantum memory for light.

\subsubsection{Quantum memory criteria}

The criteria for an IdQM (Eq.\ref{eq:crit_idqm}) and for a DMQM
(Eq.\ref{eq:crit_dmqm}) are expressed as a function of the gains
$G_{Q,P}$ and added noises $N_{Q,P}$. In the case studied here,
one has
\begin{eqnarray}
G_Q &=& \kappa^2(1- \varepsilon_a)\\
G_P & = & \frac{1-\varepsilon_a}{\kappa^2}
\end{eqnarray}
for the gains since the sign in Eq. \ref{eq:pamem} is not relevant.
One also has
\begin{eqnarray}
N_Q &=& 1 \\
N_P &=& \varepsilon_a + \frac{\varepsilon_p}{\kappa^2}
\end{eqnarray}
since we assume all the input noises to be uncorrelated and
shot-noise limited.

Based on these parameters, we will discuss the feasibility of an
ion based memory in Sec. \ref{sec:discussion}. We now turn our
attention to the experimental implementation that we propose,
based on a large ensemble of cold ions with a $1/2
\ensuremath{\to} 1/2$ transition efficiently coupled to light.

\section{Experimental implementation}

\subsection{Linear Paul trap}

\begin{figure}
\centerline{\includegraphics[width=.75\columnwidth]{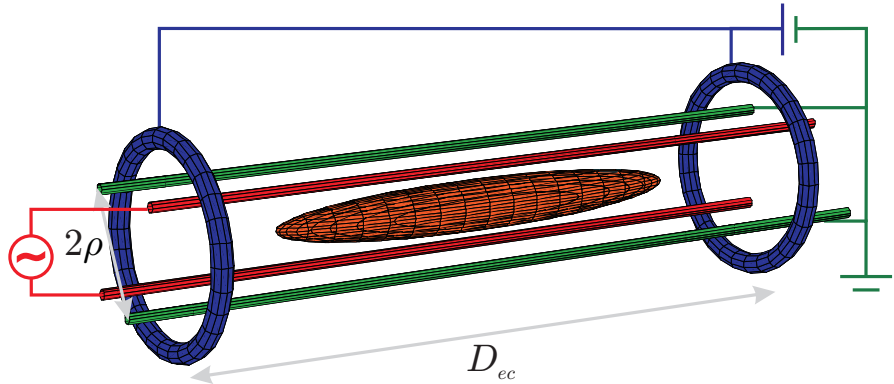}}
\caption{\label{fig:paul_trap}Principle of a linear Paul trap. Two
transverse trapping electrodes are brought to a high-frequency
high-voltage potential (in red), two are at ground voltage (green).
The longitudinal trapping electrodes are brought to a static
positive voltage (blue). The ions are trapped as an ellipso\"{\i}dal
cloud at the center of the trap (orange).}
\end{figure}

Trapping particles with external fields allows to produce well
controlled quantum systems isolated from the environment.
Furthermore, laser cooling techniques have brought the possibility
of creating dilute gases with extremely narrow velocity
distributions, thus optimizing the interaction with optical fields
and the control of the system by these fields. A standard tool to
trap and manipulate neutral atoms is the Magneto-Optical Trap
(MOT)\cite{MOT}. In such traps, the atomic sample has to be in
permanent interaction with near-resonance lasers in order to remain
trapped. This interaction induces a high rate of spontaneous
emission, leading to a rapid decoherence of both external and
internal degrees of freedom. Cold atoms based quantum-optics
experiments are thus usually carried out in the absence of the
cooling and trapping lasers on a free sample, leading to relatively
short interaction times (ms) and evolving conditions (velocity,
density...) \cite{lambrecht96}. The ion trapping techniques take
advantage of the strong electrostatic force which allows one to
tailor very deep (eV) confining potentials that are independent of
cooling lasers. The laser cooling techniques can be
applied very efficiently to the trapped sample, with a very simple
1D beam geometry that performs 3D cooling due to the strong coupling
of the motional degrees of freedom induced by the electrostatic
forces between ions. Once cooled, the ions may remain trapped and
cool in the absence of cooling lasers for long period of time (s)
compared to atom traps. A standard tool for trapping ions in quantum
optics and quantum information experiments is the Paul trap that is
based on time-varying electric potential applied to a set of
electrodes. The standard Paul trap \cite{paul90} has a
quasi-spherical geometry with few optical access. A common variant
is the linear-Paul trap \cite{raizen92} , based on the same principles but
offering more versatility in the geometry design. In order to
carry-out quantum memory experiments involving the macroscopic spin
of an ion sample, it is important to optimise the interaction of a
laser beam with the sample. Thus, we have chosen a very anisotropic
geometry, leading to quasi-1D sample, very elongated along the
propagation direction of the laser ($z$). The linear-Paul trap
(Figure  \ref{fig:paul_trap}) is made of four rods as main trapping
electrodes, two of them being brought to a high-frequency
high-voltage potential. These electrodes are responsible for a 2D
confinement in the plane ($xOy$) perpendicular to the electrodes
axis.  Two ring electrodes brought to a static positive voltage trap
the ions along the axial direction ($Oz$).

This geometry offers a wide variety of optical accesses, in
transverse direction for laser cooling and along the axis for
optimal interaction with a focused laser beam carrying the quantum
information. This set-up is currently being developed in our group.
The dimension we have chosen ($\rho=2.6$~mm, $D_{ec}=20$~mm, see
Fig. \ref{fig:paul_trap}) allows the formation of a sample of 5~mm
long, 200~$\mu$m large, containing a few $10^5$ ions \cite{coloq9}.
Using a tighter confinement, \emph{i.e.} more closely spaced
electrodes, one can expect to trap approximately 2.2~$10^6$ ions
using realistic values for the RF voltage (800~V) with a section of
60~$\mu$m and a length of 2~cm for the ions cloud.

%

\subsection{Choice of the ion}

We restrict ourselves to the second column of the periodic table
(alkaline earth elements) : these elements, when deprived of an
electron, have an electronic structure consisting of a shell with
a single orbiting electron. This structure, identical to the
alkaline elements, readily provides transitions necessary for an
easy manipulation with light fields. The available non radioactive
elements are beryllium, magnesium, calcium, strontium and baryum.
All these ions have identical electronic structures namely
2S1/2\ensuremath{\to}2P1/2 transitions. These transitions are
located from the mid (Be$^+$) to the near UV (Ba$^+$). Choosing
the last two elements allows one to use solid-state laser
technologies, such as Ti:Sa lasers. It is interesting to choose
two neighboring species (like Sr$^+$ and Ba$^+$) which can be
easily manipulated with light fields: having two species at hand
allows one to perform sympathetic cooling \cite{drewsen}. In a
sympathetic cooling scheme, the heavier element (Ba$^+$) will be
located at the perimeter of the trap and the lighter element
(Sr$^+$) at the center. This last element will thus be confined in
an elliptic-shaped region, well adapted to the interaction with a
focused laser beam. In the following, we will thus consider a
cloud of cold Sr$^+$ ions as active medium.

\section{Trapped ions for a quantum memory}

\subsection{Relevant parameters for the X-type level scheme}

As mentioned in section \ref{sec:principes}, the relevant
parameters are $\kappa$ and $\varepsilon_{a,p}$. We first compare
on Tab.\ref{tab:exp}, the values of the experimental parameters
for the proposed experiment and that used in \cite{polzik}.

\begin{table}[h]
\[
\begin{array}{|c|c|c|c|c|c|c|c|c|c|c|}
\hline
 & N_p & N_a & \begin{array}{c} \lambda\\ (nm)
\end{array} &
\begin{array}{c} \Delta \\ (MHz) \end{array}&
\begin{array}{c} \gamma \\ (MHz) \end{array}& \begin{array}{c} A
\\ (cm^2) \end{array}\\
\hline  \mbox{Ref. \cite{polzik}} & 4~10^{12} & 3
~10^{11}& 852 & 700 & 5& 6 \\
\hline \mbox{Ion cloud} & <10^{12} & 1.5~10^6 & 422 & >100 & 20 &
8~10^{-9}\\
\hline
\end{array}
\]
%
\caption{\label{tab:exp}Experimental parameters}
\end{table}

From these values, one can calculate the parameters relevant for
Eqs.~(\ref{eq:xpout}-\ref{eq:pdout}): the corresponding values of
$g$, $\kappa$ and $\varepsilon_{a,p}$ are shown in
Tab.~\ref{tab:param}. The calculation of these parameters for
experiment \cite{polzik} is more involved since the transition which
is used is $F=4, m_F=\pm 4 \ensuremath{\to} F'=5$. Thus we will use
the value for $\kappa$ given in Ref.\cite{polzik}. In this
experiment, the number of photons and atoms involved being large,
the noises induced by $\varepsilon_{a,p}$ are small (the value given
by our model are indicated in Tab.~\ref{tab:param} as a reference).
The exact values of $\varepsilon_{p}$ is not relevant since the
optical losses due to the cell windows are predominant.

\begin{table}[h]
\[
\begin{array}{|c|c|c|c|c|}\hline
\ & g^2 & \quad \kappa \quad & \varepsilon_a & \varepsilon_p \\
\hline  \mbox{Ref. \cite{polzik}} & 34.5~10^{3} & 0.37 & 5~10^{-3} &
3.5~10^{-4}\\\hline \mbox{Ion cloud} & 1.8~10^9 & 0.64 & 0.09 &
1.4~10^{-8}\\ \hline
\end{array}
\]
\caption{\label{tab:param}Relevant parameters for light-matter
interface. For the ion experiment, we have taken $N_p=2.1~10^{12}$
(corresponding to the same power as Ref.\cite{polzik}),
$N_a=1.5~10^6$, $\Delta = 8~10^3 \gamma$, $A = 1.1~10^{-8}~m^2$
(corresponding to a beam waist of 60~$\mu m$, adjusted to the
maximal cloud section).}
\end{table}

One remarks that, even though the values of $\kappa$ are comparable,
the photonic noise term, due to $\varepsilon_{p}$, is much larger in
the ion experiment. This is due to the fact that the number of
interacting atoms, $N_a$, is much smaller in this experiment. The
use of an immobile ensemble of ions allows one to choose a much
tighter focusing, increasing the atom-photon coupling constant, $g$
and keeping the value of $\kappa$ large. Furthermore, let us note
that the value of $\varepsilon_{a}$ remains very small, on the order
of a few percent, which is comparable to the other losses which will
appear in a typical quantum optics experiments (detection losses in
particular). One also remarks that $\varepsilon_a$ and $\kappa$ both
depend on the ratio $N_p/\Delta^2$. When this quantity is kept
constant, $\varepsilon_a$ and $\kappa$ will both remain constant
(provided the other parameters do not change). One can thus use a
much smaller number of photons, for instance using shorter pulses
($\mu$s), together with a smaller detuning. This approach remains
valid as long as $N_p$ remains larger than $N_a$: when $N_a\gg N_p$,
the dominant noise is the photonic noise corresponding to
$\varepsilon_p$.

\subsection{Performances of the memory, decoherence}

Let us now discuss the behavior of the proposed ion memory with
respect to the decoherence and other phenomena that may affect its
performances. We can distinguish three different phases in which a
real memory may not follow an ideal behavior: the writing phase, the
storage phase and the readout phase. We discussed in section
\ref{sec:cvqm} the effect of the losses $\varepsilon$ associated to
the writing and (eventually) to the readout phases. In particular,
in the case of a DMQM the condition $\varepsilon<1/2$ must be
fulfilled, where we consider $\varepsilon$ as the cumulative losses.
Concerning the volatility of the memory, the decoherence of the
collective spin state is the main limitation to long storage times.
Such a decoherence can be described in terms of an additional loss
$\varepsilon_{sto}$ that takes into account the spins flips or
atomic losses that may occur in the cloud. In the neutral-atom based
memory \cite{polzik} the storage time is limited by three major
phenomena: the spin-flips that can occur during atomic collisions,
the finite time spent by an atom in the laser beam, and the
dephasing induced by magnetic field inhomogeneities. The first two
phenomena are expected to give a negligible contribution in the case
of cold trapped ions. Indeed, due to the electrostatic repulsion,
the ions can not collide to give a spin-flip (core collisions): in a
raw evaluation the spin-spin and spin-orbit interaction at the
characteristic distance of $\sim 1 \mu$m  can only affect the
energy levels with a shift of the order of 100~Hz \cite{fontana62}.
Furthermore, the ion cloud being trapped by electrostatic fields
that confine strongly the ion cloud and which are not switched-off
during the experiment, no ion will move out of the laser beam during
the whole interaction time.

The main effect of magnetic field inhomogeneities is to induce a
spread in the precession frequencies around the external magnetic
field that imposes the quantification axis. Such an effect would
seem to affect an ion cloud in the same way it affects a neutral
atomic cloud. However, in an ion Coulomb crystal the position of
every ion is fixed in space. In this situation, when the readout
time for the information stored in the quantum memory is a known
parameter, it is possible to wash-out the effect of inhomogeneities
with a spin-echo experiment. Moreover, let us remark that the space
region in which the inhomogeneities have to be controlled is smaller
in the case of trapped ions: it is therefore easier to use passive
or active compensation coils to reduce the stray fields down to some
tenths of nT \cite{ringot00}.

Let us now discuss the other phenomena that could deteriorate the
performances of a quantum memory in the case of trapped ion clouds.
Large ion Coulomb crystal can be obtained with characteristic
temperatures in the range of 20--100~mK \cite{hornekaer02,roth05}.
Contrary to the case of ions near the trap axis, the ions in the
external part of the crystal are subject to a non-vanishing RF field
that induces a micro-motion. In a laser-cooled sub-Kelvin crystal,
such a micro-motion, that mainly affects the radial motion of the
ion, is not the main source of heating \cite{roth05} and its effect
is mainly limited to a Doppler Shift of the optical transition (that
is important only during the writing and reading phases). This
residual motion is well described by a thermal motion with a maximum
temperature in the 100 mK range, that gives us for Sr$^+$ a rms
Doppler shift of $\sim 10$~MHz, much smaller than the detuning
$\Delta$ and smaller than the natural line-width $\gamma$. Let us
remark again that the situation of a large ion cloud described by a
collective continuous quantum variable is very different with
respect to the case of discrete motional quantum states that are
very sensitive to any residual micro-motion. Other contributions to
the decoherence could originate from the DC Stark shift induced by
trapping electric fields and from the quadrupolar shift due to
electric-field gradients. The first contribution should vanish at
the potential minima that correspond to equilibrium position in an
ion crystal, but the ions are subject to thermal motion and explore
regions with nonzero fields and gradients. Such a problem has been
carefully analyzed in the case of single ions trapped in the center
of a small trap for metrology purposes \cite{barwood04,oskay05}. In
this case the main contribution to the shift of energy levels is the
quadrupolar coupling that, for well compensated traps can be as low
as a fraction of Hz. Even in a larger trap, we still expect a
negligible contribution to decoherence from these phenomena.
Finally, another  candidate that could affect the storage time of an
ion-based memory is the presence of collisions with the neutral
background gas in the vacuum chamber. This phenomenon is the most
important heating source in large cold Coulomb crystals
\cite{roth05}: the trapped ions that collide with a neutral particle
are not ejected from the trap but their kinetic energy increases
such that they escape from their initial site. Such kind of
collisions could naturally induce a spin-flip and their rate should
be minimized in order to achieve long storage times. The expected
rates in ion Coulomb crystals in a vacuum of  $\simeq 10^{-10}$~mbar
are of the order of ${1\over \tau}=0.03$ collisions per ion per
second \cite{drewsen}. Such a rate will induce an exponential
time-dependent loss term $\varepsilon_{sto}(t)=1 - \exp(-2t/\tau)$.

\subsection{Quantum memory criteria}
\label{sec:discussion}

As shown previously, the quality of the quantum memory can not be
evaluated independently of its purpose. It is thus necessary to
calculate the parameters $G = \sqrt{G_Q G_P}$ and $N= \sqrt{N_Q
N_P}$. These parameters can be calculated taking into account
additional optical losses before and after the interaction as well
as the decay induced by the main cause of decoherence, \emph{i.e.}
collisions with neutral particles as mentioned above.

Taking $\Delta = 3~10^4 \gamma$ and no losses, one gets $N=0.08$ and
$G=1$ without losses. If one takes into account the losses, $G$ is
now 0.96 and $N$ is 1.6. The criteria for a generalized IdQM ($G =
1$, $N<2$) is thus fulfilled for $N$ but not for $G$ except in the
absence of losses. This system is thus not suitable for such an
application.

However, it forms a very efficient DMQM, as we will show. We
calculate now the values of $G$ and $N$ in different cases (Tab.
\ref{tab:criteres}).

\begin{table}[h]
\[
\begin{array}{|c|c|c|c|c|}
\hline
\ & \quad G  \quad &\quad  N  \quad &\quad  \displaystyle \frac N G \quad \\
\hline
\begin{array}{c}
\mathrm{Ion cloud,}\\ \mathrm{no losses}
\end{array} & 0.90 & 0.31 & 0.33\\
\hline
\begin{array}{c}
\mathrm{Ion cloud,}\\ \mathrm{losses},~ t=0~s
\end{array} & 0.88 & 0.52 & 0.60\\
\hline
\begin{array}{c}
\mathrm{Ion cloud,}\\ \mathrm{losses},~t=9~s
\end{array} & 0.67 & 0.67 & 1.0\\\hline
\mbox{Ref. \cite{polzik}} & 0.84 & 0.67 & 0.80 \\
\hline
\end{array}
\]
\caption{Quantum memory parameters. The optical losses in the ion
experiment are taken to be equal to 1\% before the interaction and
5\% after, reflecting the detection losses. In experiment
\cite{polzik}, we take the losses to be evenly distributed before
and after the interaction and corresponding to the 8 uncoated
optical windows present in the experiment.\label{tab:criteres}}
\end{table}

This table shows $N/G$ is small in both cases. We remark that the
value expected for the ion cloud is smaller with respect to the
value obtained for Ref.\cite{polzik}. We also show that the storage
time during which the memory keeps its quantum character is on the
order of 9~s, very long compared to the storage times previously
demonstrated.

Let us also mention the use of ion clouds in the $\Lambda$-type
scheme performed inside a cavity \cite{pinard}. The relevant
parameter for such an experiment is the so-called cooperativity
parameter $C$ defined as
\begin{equation}
C = \frac{2 \pi g^2 N_a}{c \gamma T}
\end{equation}
where $T$ is the coupling mirror intensity transmission and the
other parameters have been previously defined. Using a typical value
$T=0.1$ for the coupling mirror transmission, one obtains for our
parameters $C \approx 55$ which is similar to the value typically
obtained with cold neutral atoms in a magneto-optical trap.

Our analysis demonstrate that, independently of the chosen
interaction scheme, a quantum memory based on a cold ion cloud is
feasible. Moreover, this approach leaves open for the
experimentalist the possibility to explore a wide region in the
parameter space, still obtaining a functional memory.

\section{Conclusion}

In this paper we presented the theoretical and practical issues
related to the realization of a quantum memory that stores quantum
information carried by continuous variables in the collective spin
observables of a large cloud of trapped and cooled ions. In
particular we discussed the different criteria that quantum
memories may have to fulfill and we presented the physical basis
of quantum recording, based on previous theoretical analysis
\cite{duan00}. By comparing the relevant physical parameters that
can be obtained using a large ion crystal to that of the unique
experimental example of an atomic quantum memory \cite{polzik}, we
demonstrated that the use of laser cooled ions (e.g. $^{88}$Sr$^+$
ions) is, under several aspects, an ideal case. We discussed the
phenomena that affect the performances of this quantum memory and
we concluded that a ion Coulomb-crystal based memory is feasible
and that it will probably outperform the neutral-atom based
quantum memories.

\section*{acknowledgement}
Laboratoire Mat\'{e}riaux et Ph\'{e}nom\`{e}nes Quantiques of the Universit\'{e}
Denis Diderot is associated with the Centre National de la
Recherche Scientifique, UMR CNRS 7162.

We acknowledge fruitful discussions with M. Pinard.

\end{document}